\documentclass[aps,prl,twocolumn,showpacs,superscriptaddress]{revtex4}
\usepackage{graphicx}
\begin{document}

\title{
Density-matrix renormalization group study of pairing when
electron-electron and electron-phonon interactions coexist: 
Effect of the electronic band structure}

\author{Masaki Tezuka}
\affiliation{Department of Physics, University of Tokyo,
Hongo, Tokyo 113-0033, Japan}
\author{Ryotaro Arita}
\altaffiliation{On leave of absence from 
Department of Physics, University of Tokyo,
Hongo, Tokyo 113-0033, Japan}
\affiliation{Max-Planck-Institut f\"{u}r 
Festk\"{o}rperforschung, Stuttgart 70569, Germany}
\author{Hideo Aoki}
\affiliation{Department of Physics, University of Tokyo,
Hongo, Tokyo 113-0033, Japan}

\date{\today}

\begin{abstract}
Density-matrix renormalization group is used to study the pairing 
when both of electron-electron and electron-phonon 
interactions are strong in the Holstein-Hubbard model at half-filling 
in a region intermediate between the adiabatic (Migdal's) and 
antiadiabatic limits.  We have found: 
(i) the pairing correlation obtained for a one-dimensional system 
is nearly degenerate with the CDW correlation 
in a region where the phonon-induced attraction 
is comparable with the electron-electron repulsion, but 
(ii) pairing becomes dominant when we destroy the electron-hole 
symmetry in a trestle lattice.  
This provides an instance in which pairing can arise, 
in a lattice-structure dependent manner, 
from coexisting electron-electron and electron-phonon interactions.
\end{abstract}
\maketitle
{\it Introduction} ---
The problem of what happens 
when the electron-electron interaction {\it coexists} 
with electron-phonon interaction 
is arousing interests from various viewpoints
\cite{HuangHankeArrigoniScalapino03, CaponeFabrizioCastellaniTosatti02,
HanKochGunnarsson00,FreericksJarrell95,
JeonParkHanLeeChoi04}.  
In fact, there are a number of classes of materials 
where electron-electron (el-el) and electron-phonon (el-ph)
interactions are both significant.  One example is 
the superconducting doped fullerenes\cite{Gunnarsson}, 
where the electrons in narrow conduction bands 
are strongly coupled to intra-molecular, high-frequency phonons.  
Theoretically, a fundamental question is: can superconductivity arise 
in the coexistence of 
the el-el and el-ph interactions? It may seem difficult 
for them to work constructively, 
since the effective electron-electron attraction arising from 
the el-ph coupling favors isotropic pairs, while 
the (spin-fluctuation mediated) pairing interaction 
due to the el-el repulsion favors 
anisotropic pairs.  The problem becomes especially 
nontrivial for a region intermediate between the adiabatic (Migdal's) and 
antiadiabatic limits.  

There is a theoretical reason why we have to look into the 
intermediate regime. 
Superconductivity (SC) has to compete with diagonal orders 
in general, and SC phases in fact often 
arise adjacent to density-wave phases on the phase diagram. 
In the present context, a strong el-ph interaction will favor 
a charge density wave (CDW), 
while a strong el-el interaction will favor a spin density wave (SDW).   
A metallic phase has a chance to appear around 
the boundary where CDW gives way to SDW, at which the el-el and el-ph 
interactions are indeed comparable.  
Takada and coworkers have in fact argued 
that an off-site pairing arises in a region between CDW and SDW phases, 
from the enhancement factors in the CDW, SDW and SC response 
functions in a two-site result by assuming that the 
two-site system is already close to the infinite system\cite{Takada95a}, 
and from an exact diagonalization for two- and four-site 
Holstein-Hubbard model at half-filling\cite{HottaTakada97}.  
However, a study for larger systems may be desirable 
to elaborate why and how the pairing occurs when the el-el 
and el-ph interactions 
are comparable.
This region is physically interesting 
as the case of interacting fermions and bosons with similar energy scales,  
but is technically challenging as well, since we are not allowed to 
adopt the Migdal's approximation 
for the phonon energy $\hbar \omega$ assumed to be much smaller
than the electron energy scale, $t$, nor can we adopt the 
antiadiabatic limit ($\hbar \omega \gg t$).  

The present Letter focuses on this problem, where we adopt 
the density-matrix renormalization group (DMRG)\cite{White} 
for a one-dimensional Holstein-Hubbard model.  
DMRG can treat interactions non-perturbatively,
and various types of correlation functions at $T=0$ can be 
obtained from the wavefunctions.
A crucial interest here is how the system is 
dissimilar (or similar) to purely electronic systems, 
so we scan the three parameters (the three axes in the 
inset of Fig.1) characterizing the 
system: $U/t$ (the electron-electron 
repulsion $U$ divided by $t$), 
$\hbar\omega/t$ (ratio of the phonon and electron energy scales), 
and $\lambda \equiv 2g^2/\hbar\omega$ 
(the effective attraction between electrons on the same site 
as defined in the $\omega\rightarrow\infty$ limit).
As for the electronic band filling, we have concentrated here 
on the half-filling as in 
\cite{Takada-Chatterjee,Hager,Fehske,JeonParkHanLeeChoi04}, 
having $\mbox{A}_{3}\mbox{C}_{60} (\mbox{A}=\mbox{K},\mbox{Rb})$ in mind.

We shall conclude that, 
(i) while the correlation functions obtained here indicate that 
superconductivity does not dominate over, but 
is nearly degenerate with the CDW correlation 
(a curious similarity with the behavior in 
purely electronic systems where the two correlations 
should coincide with each other when the system 
is electron-hole symmetric), but 
(ii) pairing becomes dominant when we destroy the electron-hole 
symmetry (in a trestle lattice).  
These occur in the region of interest, 
where the phonon-induced attraction 
almost cancels the electron-electron repulsion, and 
intermediate between the adiabatic and antiadiabatic limits.  
So the message here is 
that the coexistence of el-el and el-ph 
interactions can, {\it in a manner dependent on the underlying electronic 
structure}, give rise to pairing.

{\it Method ---} 
Inclusion of the phonon degrees of freedom makes the DMRG 
calculation more demanding.
However, Jeckelmann and White 
have introduced the pseudo-site method\cite{Jeckelmann-White}, 
which makes the application of DMRG to
models with on-site (Einstein) phonons feasible, and 
charge and spin gaps have been calculated 
for the one-dimensional Holstein-Hubbard model\cite{Hager, Fehske}, 
\begin{eqnarray}
 H&=&-t\sum_{i,\sigma}(c_{i+1,\sigma}^\dag c_{i,\sigma}+{\rm H.c.})
+U\sum_i n_{i\uparrow}n_{i\downarrow}\nonumber\\
&+&g\sum_{i,\sigma}n_{i\sigma}(b_i+b_i^\dag)
+\hbar \omega\sum_i b_i^\dag b_i.
\label{eqn:HH}
\end{eqnarray}
Here $c_{i\sigma}^\dag$ creates an electron of spin $\sigma$ on the
$i$th lattice site, $t$ is the nearest-neighbor hopping 
(which we take as the unit of energy hereafter),
$U$ is the on-site electron-electron repulsion,
$a_i^\dag$ creates a phonon at $i$th, 
$g$ is the on-site el-ph interaction, 
and $\hbar \omega$ is the bare phonon energy.  
In the pseudo-site method, 
we consider only finite numbers of phonons at each site 
($M=2^N$ for 0-boson, 1-boson, \ldots, $2^{N}-1$-boson states).  
Sites are expressed 
in terms of fictitious $N$ phonon ``pseudo-sites'',
on top of an electron pseudo-site.   
The pseudo-sites are 
taken into the system one by one, as in real sites in conventional DMRG.  
Thus the maximum dimension of the Hilbert space considered in the
DMRG depends only on the maximum number of retained states $m$
and not on $M$.  We have here retained 
up to $m=600$ states in each block of the DMRG
calculation for each chain with $L=40$ sites with $M=2^4$ states 
considered for each site. 

In calculating the correlation functions, we have noticed a flaw 
in the conventional infinite-size DMRG algorithm, in which two new sites 
are inserted between two symmetrical blocks.  
For the Holstein-Hubbard model, 
the retained states when the electron pseudo-site is added
tend to significantly deviate from those describing the system 
at later stages.  
This is because the electrons on the added pseudo-sites 
feel the bare repulsion there, while the repulsion 
on other sites is reduced due to the electron-phonon coupling, 
and this degrades the convergence.  
So we propose\cite{physica} 
to remedy this by modifying the chemical potential for the
electrons for the added electron pseudo-sites 
in the calculation for the ground states, 
so that the expectation value of the number of electrons at
each of the new pseudo-sites equals those 
in the target state, which is unity 
for the half-filled band assumed here.
This method, which we call the compensation method, 
has indeed given lower ground state energies and hence 
more accurate ground states and correlation functions.

\begin{figure}
\includegraphics[width=8cm]{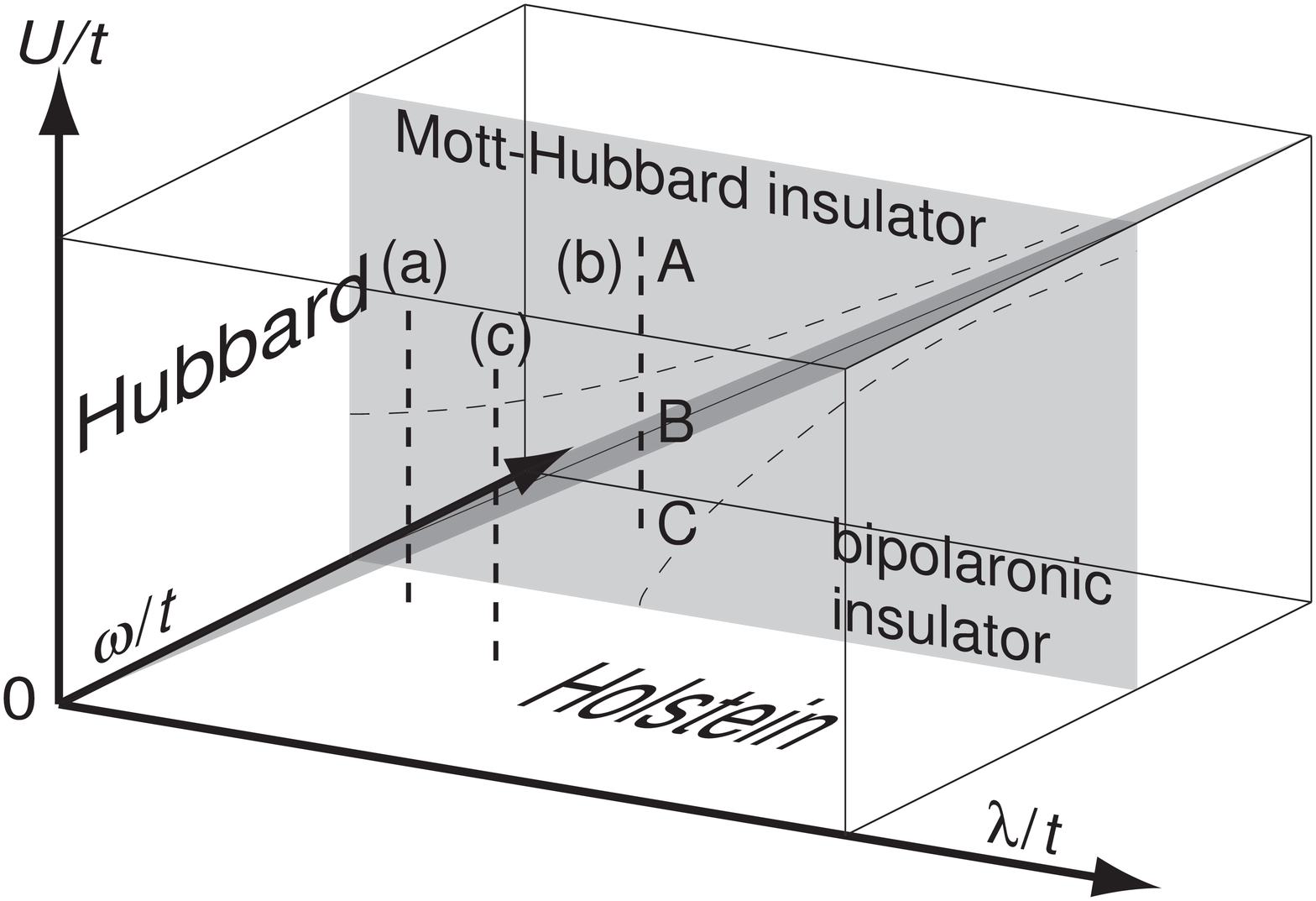}\\
\includegraphics[width=8cm]{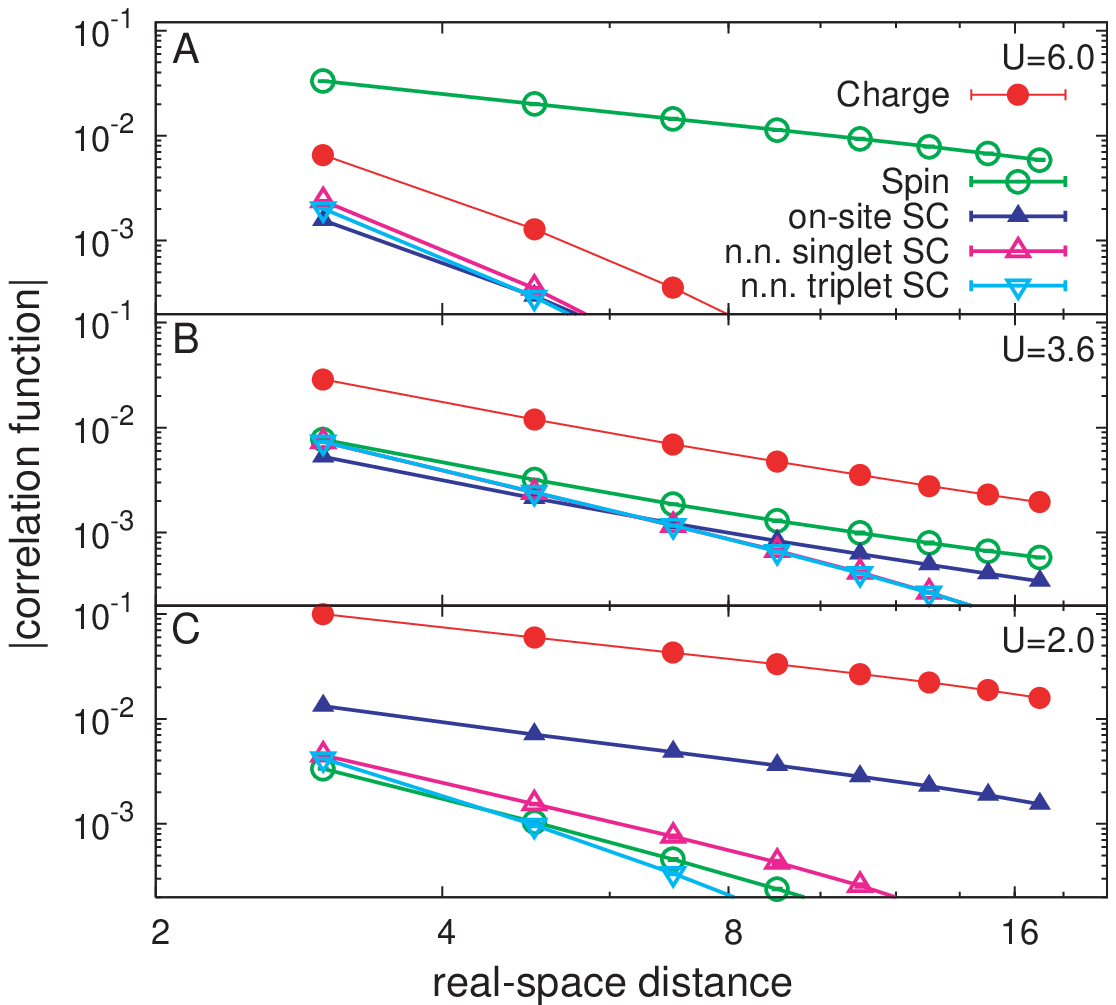}
\caption{
Various correlation functions versus real-space distance 
in the Holstein-Hubbard model at half-filling calculated for 
$(\lambda,\omega)=(3.6, 5.0)$ (in units where $t=1, \hbar =1$) 
with three values of $U$, 
indicated as A,B,C in a schematic parameter space (top inset). 
Error bars are smaller than the size of each symbol.  
Only the data for $|i-j|=$odd are plotted,
because of an even-odd effect in correlation functions 
in open-boundary condition. 
Vertical arrows with (a),(b) correspond to
the scans in Fig.~\ref{fig:power}(a)(b).
\label{fig:correlation}}
\end{figure}

{\it Results} ---
We have calculated the correlation function 
$\langle O_i^\dag O_j\rangle$ for
$O_i=n_{i\uparrow}+n_{i\downarrow}$ (charge),
$(n_{i\uparrow}-n_{i\downarrow})/2$ (spin),
$c_{i\uparrow}c_{i\downarrow}$ (on-site pair),
$c_{i\uparrow}c_{i+1\downarrow}-c_{i+1\uparrow}c_{i\downarrow}$
(nearest-neighbor spin-singlet), and
$c_{i\uparrow}c_{i+1\downarrow}+c_{i+1\uparrow}c_{i\downarrow}$ 
(nearest-neighbor triplet).  Typical result is displayed, 
on double-logarithmic scales, 
in Fig.\ref{fig:correlation} 
for the cases where (A) $U>\lambda$, 
(B) $U=\lambda$, and (C) $U<\lambda$.  
After the infinite-algorithm DMRG calculation is performed 
with the compensation method, 
at least three sweeps of the finite-algorithm are done.  

We immediately notice the following: 
if we concentrate on power-law correlations, 
they are the charge and on-site pair correlations 
for $U\lesssim\lambda$, or the spin correlation for $U\gtrsim\lambda$, 
while all of the charge, spin and on-site pair correlations 
decay with power laws for $U\simeq\lambda$.  
This is our first key result.
It is interesting to compare the 
present correlation functions with 
an estimation of spin and charge gaps in the same system by 
Fehske {\it et al}.\cite{Fehske}
They suggested that a charge gap opens and spin is gapless
when $U\gg \lambda$, 
both gapless when $U\simeq \lambda$, 
both gapful when $U\ll \lambda$.  The former two agree with the 
present result.
As for the latter, we can interpret that 
here we are looking at a region
where $U$ is only moderately smaller than $\lambda$
and a spin-gapped metal can exist,
for which CDW and on-site SC have power-law correlations
as far as the Tomonaga-Luttinger physics\cite{TLref} goes.  

\begin{figure}
\includegraphics[width=8cm]{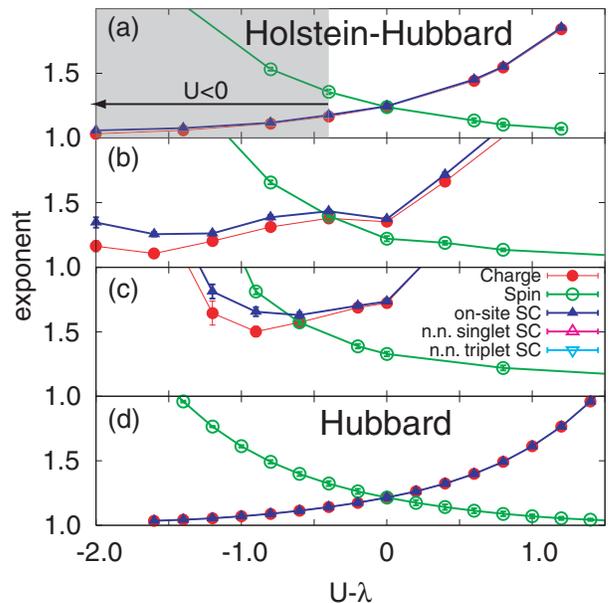}
\caption{Calculated powers ($\eta$) for the 
correlation functions ($\propto r^{-\eta}$) 
against $U-\lambda$ for 
(a) $(\lambda,\omega)=(0.4,5.0)$, 
(b) $(\lambda,\omega)=(3.6,5.0)$ and
(c) $(\lambda,\omega)=(3.6,2.5)$
in an $L=40$, half-filled 
Holstein-Hubbard chain,
where $U$ is varied along the dashed arrows in inset of Fig.1.
Gray shadow indicates $U<0$.
 (d) The result for the Hubbard model (with $\lambda=0$), where the curves for 
charge and on-site SC are degenerate.  
\label{fig:power}}
\end{figure}

So a next key issue is whether the exponents are 
solely determined by $U-\lambda$, which would be 
the case if the TL picture somehow holds everywhere.  
We have looked at its systematic dependence, 
and Fig.~\ref{fig:power} plots typical behaviors.  
For a smaller $\lambda$ ($\propto g^2$) 
the dominant correlation in Fig.~\ref{fig:power}(a) 
changes from the CDW/on-site SC 
to SDW almost exactly at the point where $U$ exceeds $\lambda$.

To be more precise, 
it is difficult, for finite systems, to distinguish whether 
the decay of a correlation function is power or exponential.  
So we have checked how the result compares with the 
exponent, calculated in the same way, 
for a purely electronic (i.e., the Hubbard) model 
for the same system size in Fig.~\ref{fig:power}(d).  
In the one-dimensional Hubbard model, the correlations 
that decay with a power law ($\sim 1/r$) are 
on-site SC and CDW for attractive $U<0$ (SDW for repulsive $U>0$).  
In Fig.~\ref{fig:power}(d), this appears as a crossing of 
two, continuous curves (one of which are doubly-degenerate) at $U=0$, 
and the result resembles  Fig.~\ref{fig:power}(a). 
For finite systems we have also to be careful about the effect of 
discreteness of the 
levels (against spin and charge gaps) on the correlation 
function\cite{KurokiKimuraAoki96}.  
We have checked that the spin gap for $(U,\lambda,\omega)=(2.0,3.6,5.0)$ 
extrapolated from the $L\leq50$ result 
is about twice as large as the difference between the lowest-unoccupied and
highest-occupied one-electron levels.
So finite-size effects would not be significant, at least for
$|U-\lambda| \gtrsim 1.0$.

However, we have found that, when $\lambda$ is increased 
(with a fixed adiabaticity $\omega/t$), 
the crossing point does deviate from 
$U-\lambda=0$ as seen in Fig.~\ref{fig:power}(b).  
Also, the degeneracy between CDW and on-site SC is 
lifted (with the power for the latter slightly larger).
This becomes more manifest for smaller $\omega$,
as shown in Fig.~\ref{fig:power}(c) for $\omega=2.5$.  
There again, CDW and on-site SC correlations both have power-decay, 
but they start to decay exponentially 
when $U$ is further decreased, signalling an opening of the 
charge gap $\Delta_c$ as well as spin gap $\Delta_s$.  
This should correspond to the region of  $\Delta_c \sim \Delta_s >0$ 
for $U \ll \lambda$ in Ref.\cite{Fehske}.  
So, as we move away from the $\omega\rightarrow\infty$ limit where
the spin-gapped metallic region appears for $U<\lambda$, 
the region becomes bounded from both below 
(opening of $\Delta_c$) and above (CDW-SDW crossing point).  
The decrease in $U-\lambda$ at the CDW-SDW crossing 
may be related to the effective repulsion enhanced from $U-\lambda$ 
as studied in \cite{SangiovanniCaponeCastellaniGrilli05}
with the dynamical mean field theory (DMFT).

If we assume, as in \cite{ClayHardikar05},
that the Tomonaga-Luttinger (TL) theory\cite{TLref} applies to 
the present system, in a spin-gapped metal the CDW correlation should
behave as $r^{-K_\rho}$ while the pairing correlation $r^{-1/K_\rho}$, 
so that $K_\rho>1$ corresponds to dominant SC correlation.  
While the behavior of the CDW correlation with a power greater than unity 
is consistent with the results in \cite{ClayHardikar05},
we also notice that the power for the on-site SC is larger 
than that for CDW, which would not occur if the TL description persisted.  
Since this is observed over the entire range of $U-\lambda$, 
this should not be due to a finite-size effect from the above argument.  
So, while a deviation from the TL does exist 
(which is not surprising since the coupling to phonons 
should make the el-el interaction effectively energy-dependent), 
the on-site SC does not become the most dominant 
correlation in the region studied here. 
We note that this robustness of CDW against SC at half-filled Holstein-Hubbard
model has also been observed in a DMFT calculation for the transition temperature
\cite{FreericksJarrell95}.

About the curious near-degeneracy between the 
CDW and on-site SC correlations, we can say the following.  
If the system were purely electronic, 
then an electron-hole transformation (for down spins) 
maps, when the lattice is bipartite, 
the CDW correlation onto the SDW correlation ($\langle S^zS^z\rangle$) 
and the on-site SC correlation onto the SDW correlation 
($\langle S^-S^+\rangle$, which should be identical with 
$\langle S^zS^z\rangle$ due to the SU(2) symmetry).  
When the Hamiltonian is the (half-filled) Hubbard model, the 
electron-hole transformation is the well-known attraction-repulsion 
($U\leftrightarrow -U$) transformation,\cite{nagaoka} and 
the CDW and on-site pairing correlations are genuinely degenerate 
(as displayed in Fig.\ref{fig:power}(c)).  
Now, the system at hand is a coupled electron-phonon system, 
which can be mapped to the Hubbard 
model only in the limit of $\omega\rightarrow\infty$.  
So the present result amounts that the degeneracy curiously 
persists approximately for the Holstein-Hubbard model 
for finite values of $\omega$. 

\begin{figure}[h]
\includegraphics[width=8cm]{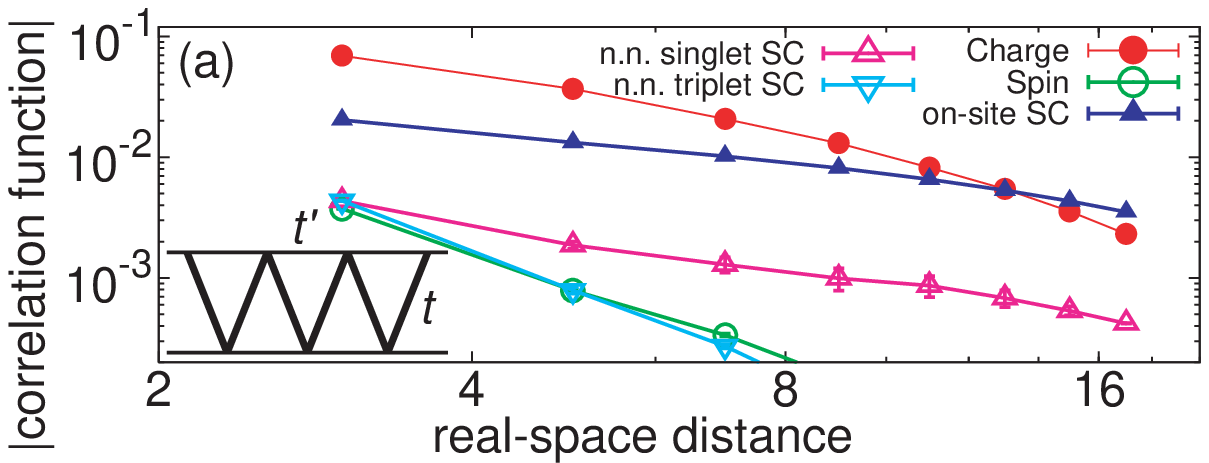}\\
\includegraphics[width=8cm]{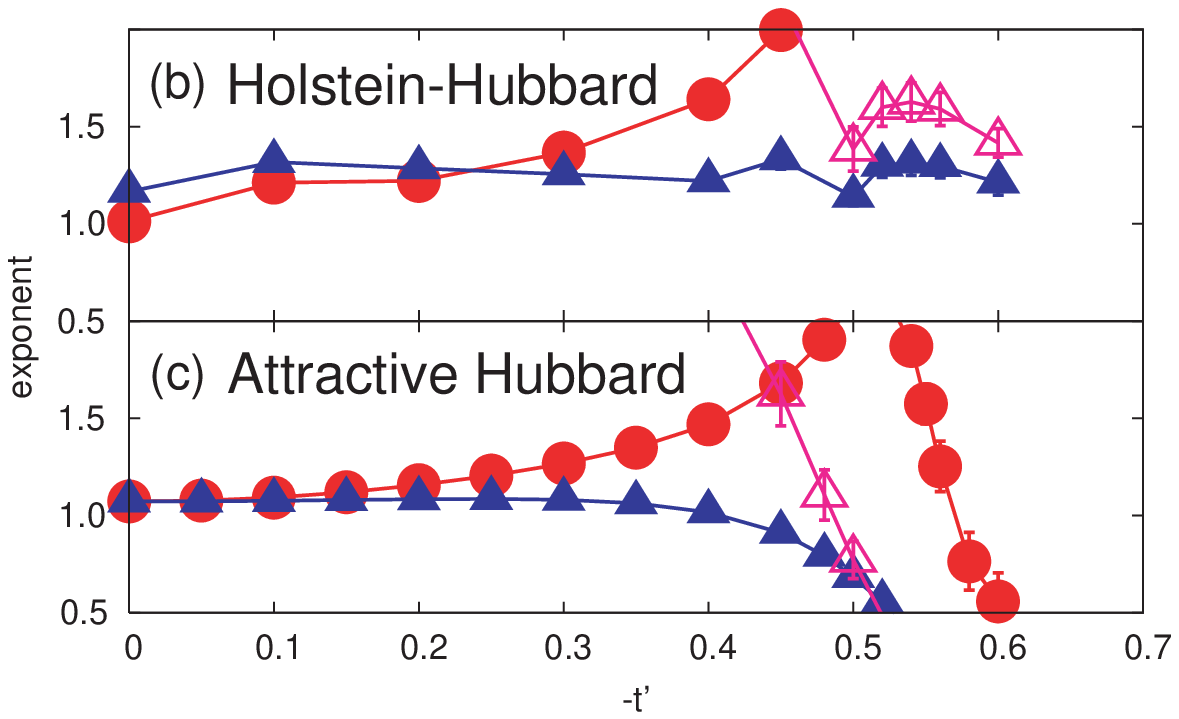}
\caption{(a) Correlation functions when the electron-hole symmetry 
is destroyed with $t'=-0.5$ for $(U,\lambda,\omega)=(2.0, 3.6, 5.0)$.
(b) Exponents against $-t'$ for various 
correlation functions for $(U,\lambda,\omega)=(2.0,3.6,5.0)$, and 
(c) a similar plot for the attractive Hubbard model with $U=-1.6$.  
The inset in (b) depicts the trestle lattice. 
\label{fig:onsite}}
\end{figure}

{\it Destruction of the electron-hole symmetry} --- 
This leads us to the following idea: 
a destruction of the electron-hole symmetry in the (one-electron) band 
structure with e.g. an introduction of the second-neighbor hopping $t'$ 
may possibly be a way to make 
the pairing correlation the single, 
most dominant correlation, as in the on-site pair correlation 
in the half-filled attractive $t$-$t'$-$U$ model.  
So we introduce the second-neighbor hopping term, 
$-t'\sum_{i,\sigma}(c_{i+2,\sigma}^\dag c_{i,\sigma}+{\rm H.c.})$ 
($\equiv$ trestle lattice; see inset of Fig.\ref{fig:onsite}), 
in the Holstein-Hubbard model.

The result in Fig.\ref{fig:onsite}(a), 
for the same $(U,\lambda,\omega)=(2.0, 3.6, 5.0)$ 
as in Fig.~\ref{fig:power}(b), shows that 
the introduction of $t'$ makes the CDW 
correlation suppressed (i.e., the power increases), 
while the power for the on-site SC correlation does not 
significantly change.  
This persists, as shown in Fig.\ref{fig:onsite}(b) 
plotting various exponents against 
$t'$, up to $t'=-0.5$, at which the number of Fermi points 
in the non-interacting band increases from two.   
So in the region $0.25 \lesssim -t' \lesssim 0.5$ 
the on-site SC becomes the single, most dominant correlation.
We also note that the nearest-neighbor singlet pair correlation, too, 
becomes dominant (i.e., decays 
more slowly than the charge, spin or nearest-neighbor triplet pair
correlations), and becomes nearly as dominant as the on-site pair 
when $-t'$ approaches $0.5$.  
Here again, 
the calculated exponents for the Holstein-Hubbard model 
violates the TL relation, $\eta_{CDW}\eta_{SC}=1$, 
for a spin-gapped metal.

{\it Discussion} ---
We have obtained 
an instance where superconductivity appears in a manner 
dependent on the lattice structure, while one might think that 
the electronic band structure should be irrelevant 
for the on-site (an s-wave) pairing arising from the 
on-site $U$ and on-site phonon.  
The result, if extensible to general dimensions, is suggestive 
for superconducting alkali-fullerides
$\mbox{A}_{3}\mbox{C}_{60}$
having an fcc (i.e., electron-hole {\it a}symmetric) array of fullerenes, 
while DMFT studies for superconductivity primarily assume 
a symmetric (semielliptic) electron density of states.  
Non-half-filled bands are also interesting, and the study is under way.  

\begin{acknowledgments}
This work is in part supported by a Grant-in-Aid for Science Research on
Priority Area from the Japanese Ministry of Education.
\end{acknowledgments}

\end{document}